# An appeal for an open scientific debate about the proximal origin of SARS-CoV-2

Jacques van Helden[1,2,*], Colin D Butler[3], Bruno Canard[4], Guillaume Achaz[5,6], François Graner[7], Rossana Segreto[8], Yuri Deigin[9], Fabien Colombo[10], Serge Morand[11], Didier Casane[12,13], Dan Sirotkin[14], Karl Sirotkin[14], Etienne Decroly,[4,*]  José Halloy[15,*]

1. CNRS, Institut Français de Bioinformatique, IFB-core, UMS 3601, Evry, France
2. Aix-Marseille Univ, Inserm, laboratoire Theory and approaches of genome complexity (TAGC), Marseille, France; ORCID: 0000-0002-8799-8584
3. National Centre for Epidemiology and Population Health, Australian National University, Canberra, Australia
4. Aix-Marseille Univ, CNRS, UMR 7257, AFMB, Case 925, 163 Avenue de Luminy, 13288 Marseille Cedex 09, France. ORCID: 0000-0002-6046-024X
5. Eco-Anthropologie (UMR7206 Université de Paris-CNRS-MNHN), Muséum National d'Histoire Naturelle, Paris, France
6. Center for Interdisciplinary Research in Biology (UMR7142 Collège de France-CNRS-INSERM), Collège de France, Paris, France
7. MSC, Université de Paris, CNRS UMR 7057, 10 rue Alice Domon et Léonie Duquet, 75205 Paris Cedex 13, France
8. Department of Microbiology, University of Innsbruck, Austria
9. Youthereum Genetics Inc., Toronto, Canada.
10. Université Bordeaux Montaigne, Mediation, Information, Communication, Art (MICA, EA 4426), 10 esplanade des Antilles, Pessac, France.
11. Montpellier Université, CNRS, Institut des Sciences de l'Évolution (ISEM), 34290 Montpellier, France, ORCID: 0000-0003-3986-7659
12. Université Paris-Saclay, CNRS, IRD, UMR Évolution, Génomes, Comportement et Écologie, 91198, Gif-sur-Yvette, France
13. Université de Paris, UFR Sciences du Vivant, F-75013 Paris, France
14. Karl Sirotkin LLC. ORCID: 0000-0002-9685-0338
15. Université de Paris, LIED, CNRS UMR 8236, 85 bd Saint-Germain, 75006 Paris, France. ORCID: 0000-0003-1555-2484

 * Corresponding authors

**Abstract**

One year after the onset of the COVID-19 pandemic, the origin of SARS-CoV-2 still eludes humanity. Early publications firmly stated that the virus was of natural origin, and the possibility that the virus might have escaped from a lab was discarded in most subsequent publications. However, based on a re-analysis of the initial arguments, highlighted by the current knowledge about the virus, we show that the natural origin is not supported by conclusive arguments, and that a lab origin cannot be formally discarded. We call for an opening of peer-reviewed journals to a rational, evidence-based and prejudice-free evaluation of all the reasonable hypotheses about the virus' origin. We advocate that this debate should take place in the columns of renowned scientific journals, rather than being left to social media and newspapers.

**Viewpoint**

On February 19, 2020, three weeks after the publication of the SARS-CoV-2 genome,[1] twenty-seven scientists signed a *Statement in support of the scientists, public health professionals and medical professionals of China combatting COVID-19* in *The Lancet*.[2] They took an authoritative position about the origin of the novel coronavirus behind the pandemic: *"Scientists from multiple countries have published and analysed genomes of the causative agent, SARS-CoV-2, and they overwhelmingly conclude that this coronavirus originated in wildlife"*. This statement has since attracted 23,000 additional signatures, and was used throughout the international press as proof that SARS-CoV-2 emerged due to a natural zoonosis.

We share our colleagues' annoyance about various unfounded theories spreading over social networks which seemed to be aimed at increasing geopolitical tensions. However, on the basis of the current scientific literature, complemented by our own analysis of coronavirus genomes and proteins,[3–5] we hold that there is currently no compelling evidence to definitively arbitrate between a completely "natural" origin (i.e. a virus that has evolved and been transmitted to humans solely via contact with wild or farmed animals) and a "laboratory" origin (which might involve one or more steps such as transport of animal samples to Wuhan, viral evolution, an index case occurring through viral exposure in a laboratory, accidental laboratory escape, faulty autoclaving equipment, or any other possible escape pathway ...).

Of the nine references cited in the statement to support the natural origin hypothesis, eight consist of trees showing phylogenetic relationships between SARS-CoV-2 and other coronaviruses. A distinction has to be made between the general ancestry and the *proximal* origin of the virus, i.e. the last step of its transmission chain from its original animal reservoir (putatively bats) to humans. To the best of our knowledge, the fact that the causative agent of COVID-19 descends from natural viruses has not been questioned by anyone, but this distal origin does not explain how it came to be able to infect humans. This step is still unknown, since the closest animal virus at our disposal (RaTG13) shows a 4% difference with SARS-CoV-2, genetic distance which has been estimated to reflect 4 to 7 decades of evolutionary divergence[6]. Pangolins are no longer considered a plausible intermediate host based on the molecular evidence.[7–10] We thus still need to trace the animal intermediates between the bat reservoir, locate the places of transmissions, characterize the viral strains, trace back the outbreak from the first COVID-19 patients, and then finally understand the ultimate conditions of the transfer from animals to humans.

The proximal origin was explicitly addressed in one reference in the Lancet statement: a preprint by Andersen *et al.*, later published in Nature Medicine in April 2020[11]. This article was highly influential: within 9 months it was cited in 2,000 scientific publications, and the vast majority of scientists, including many of us, initially took it for granted that this novel coronavirus was of natural origin. However, upon re-analysis, we realised that a conclusive proof of the proximal origin is still lacking. The initial method of reasoning, endorsed in many subsequent papers, was to contrast two opposing possibilities: natural origin versus "laboratory construct or purposefully manufactured virus". Two main arguments were presented against the latter possibility: (i) the specific mutations that confer their particular affinity to the SARS-CoV-2 spike protein were unknown before COVID-19 emergence, and thus could not have been designed ; (ii) the SARS-CoV-2 genome has no evidence for reverse engineering (e.g. a previously used viral backbone). The lab construct hypothesis was thus rejected, leaving a natural proximal origin as the only possibility. However, this reasoning suffers from a logical fallacy. Proving a hypothesis by discarding its alternative is only valid if the two hypotheses are mutually exclusive, and cover all conceivable possibilities. In this case, these conditions are not met, since other mechanisms are plausible, for example serial passage experiments, [12] which consist of testing and measuring the ability of a virus to infect different animal models or cultured cells. Such experiments exert an artificial selection of the random mutations that increase the fitness of the virus to the new host, thereby resulting in a fast evolution of genomic sequences. As in many virology labs, passage experiments are routinely performed in the Wuhan Institute of Virology (WIV)[13–15], consistently with their mission to collect and monitor viral strains having epidemic potential into humans. Selection during passage is dismissed by Andersen, based on the argument that it would be less parsimonious than the pangolin origin. However, the pangolin hypothesis has since been abandoned. [7–10] Regarding the hypothesis of a laboratory construct, absence of evidence is not evidence of absence, and a viral genome might be engineered with a yet unpublished backbone. Also, the expectation of finding traces of engineering in the sequences does not account for the seamless technologies currently used to synthesise nucleic acids, which have been around for about 20 years[16].

Experiments involving pathogenic viruses require highly secure laboratory conditions.[17,18] There are, however, many well-documented cases of pathogen escapes from laboratories, including viruses.[12,19–22] This scenario was a priori discarded in the February statement: *"We stand together to strongly condemn conspiracy theories suggesting that COVID-19 does not have a natural origin"*. [2] We, like The Lancet authors, condemn conspiracy theories. However, an accident is not a conspiracy, and we think that scenarios involving a potential lab accident should be evaluated rigorously, along with the other hypotheses. Even more, it is precisely because actual conspiracy theories are so rapidly spreading on social media and via some politicians that we ought, as a scientific community, to evaluate all hypotheses on a rational basis. We need to weigh their likelihood, based on facts and evidence, devoid of speculation concerning alleged political intent. This approach seems consistent with the views presented near the conclusion of the Lancet statement *"to promote scientific evidence and unity over misinformation and conjecture"*, but a little word makes a whole difference: a scientific question has never been solved, and should never be approached, by asking scientists to promote unity. Science, by definition, explores and embraces alternative hypotheses, contradictory arguments, verification, refutability, and even controversy. Departing from this principle risks establishing dogmas, and abandoning science.

Unfortunately, the unitary view promoted in the Lancet statement has, to date, been widely adopted, with few exceptions.[3,4,12,23–25] Scientific evaluations of alternative hypotheses for the origin of COVID-19 are, as yet, absent from the most prominent scientific journals. This lacuna may even fuel conspiracy theories. Instead, the scientific community should bring this debate to the place it belongs: the columns of renowned scientific journals[26]. An evidence-based, independent and prejudice-free evaluation of all the reasonable origin scenarios will require collecting samples and data in all the potentially relevant places, including wildlife sites and farms (as scheduled for the WHO mission) but also in hospitals and in laboratories. This effort is crucial, not only to solve many currently unanswered questions and elucidate the cause of the current pandemic, but also to take appropriate measures of prevention.